# Network Services Anomalies in NFV: Survey, Taxonomy, and Verification Methods

Moubarak ZOURE
Univ. Bordeaux, Bordeaux INP, CNRS, LaBRI, UMR5800, F-33400 Talence, France

moubarak.zoure@u-bordeaux.fr

Toufik AHMED
Univ. Bordeaux, Bordeaux INP, CNRS, LaBRI, UMR5800, F-33400 Talence, France

tad@labri.fr

Laurent RÉVEILLÈRE
Univ. Bordeaux, Bordeaux INP, CNRS, LaBRI, UMR5800, F-33400 Talence, France

laurent.reveillere@u-bordeaux.fr

*Abstract*— Network Function Virtualization (NFV) has emerged as a disruptive networking architecture whose galloping evolution is prompting enterprises to outsource network functions to the cloud and ultimately harvest the fruits of cloud computing, including elasticity, pay-as-you-go billing model, and on-demand services provisioning.

However, many reluctant enterprises oppose the benefits of this outsourcing to their critical and pressing concerns about security, trust, and compliance. The latter anticipate possible security and QoS policy violations stemming from dishonest behaviors by cloud providers, attacks by co-resident competitors, misconfiguration by cloud administrators, or implementations flaws by NFV developers. As a result, migrating sensitive workloads to the cloud requires enterprises to first assess risks by gaining knowledge of possible network services' anomalies and second, to build trust in the cloud by designing effective mechanisms to detect such anomalies.

This survey provides scrutiny of network services anomalies that may occur in the NFV environments. We first present a taxonomy of network service anomalies and analyze their negative impacts on critical service attributes, including security and performance. Second, we compare and classify the existing anomalies' verification mechanisms from the literature. Finally, we point out the literature gap and identify future research directions for anomalies verification in NFV.

## I. INTRODUCTION

Enterprises struggle with extreme challenges to deploy, maintain, and evolve traditional networks. These physical-wired networks rely on proprietary and hardware-based middleboxes that involve significant initial investments, vendor lock-ins, rapidly changing expertise, manual and endless configuration, and large servers' rooms. Over the last decade, Network Function Virtualization (NFV) [1] has emerged as an innovative networking architecture that aims to overcome the limitations of traditional networks. NFV exploits sophisticated virtualization technologies to implement hardware-based middleboxes as software appliances, called Virtual Network Functions (VNFs), e.g., firewalls, NATs, WAN optimizers. VNFs are consolidated onto virtual machines (VMs) or containers deployed on commodity servers located in in-house data centers or any public cloud, e.g., Amazon Web Services, Microsft AZURE, Google Cloud Platform. The NFV market will grow by 34,9% each year, reaching 122 billion dollars by 2027, according to a forecast of ResearchAndMarkets [2].

By adopting NFV, enterprises, known as *tenants*, have the opportunity to outsource their on-premises middleboxes to the cloud [3]-[5]. On the one hand, they capitalize on cloud advantages, including increased availability, flexibility to try new services, on-demand services provisioning, and pay-as-you-go billing. On the other hand, tenants lose direct control and visibility over their data, security policy enforcement, and VNFs execution when moving their services to the cloud. Furthermore, tenants lose guarantees on meeting their continually evolving compliance obligations [6], e.g., GDPR, HIPAA, ISO 270001, SAS 70, PCI DSS. Violations of these obligations may severely harm tenants' business and often result in heavy fines, interminable trials, or a tarnished reputation.

Thus, despite the enormous benefits of outsourcing, enterprises express a solid and reasonable reluctance to migrate their network services to the cloud. Many enterprises still consider the cloud as an untrustworthy domain to operate mission-critical applications. They anticipate dishonest behaviors by cloud providers, including violating strategic services clauses such as data privacy and Service Level Agreement while obscuring these violations. Also, co-resident tenants, external attackers, or malicious insiders may compromise enterprises' network services to steal sensitive data or sabotage business operations. These risks are exacerbated in cloud-based NFV environments. The NFV architecture introduces a new software stack that exposes a more extensive surface attack than traditional networks. For instance, a malicious insider [8] with sufficient privilege on the cloud control points could reconfigure network services and violate enterprises' security policies. NFV software vulnerabilities described in CVE-2018-15402 and CVE-2020-3236 allow an external attacker to gain elevated privileges or access confidential data.

Thus, tenants must inexorably draw a complete picture of possible network services' anomalies before considering effective verification mechanisms to ensure their network services' compliance against their specifications. However, the literature lacks an in-depth analysis of network services' anomalies and their related-verification approaches. Previous surveys [9]-[13] focused exclusively on threats and vulnerabilities of NFV without identifying the anomalies they may introduce in the deployed network services.

This gap in the literature motivates us to conduct the current research survey and identify significant research directions. We aim to provide a comprehensive overview of network services' anomalies and the existing techniques to verify them.

**Contributions**. This survey includes the following contributions:
- First, we analyze the threats and vulnerabilities inherent in the NFV framework that may lead to network services anomalies. We then propose a generic definition for a network service anomaly.
- Second, we scrutinize possible network services' anomalies in NFV, emphasizing their detrimental impact on well-known service attributes such as security, performance, and availability. This study provides the reader with a comprehensive taxonomy of network services' anomalies in NFV environments.
- Third, we explore and compare the existing techniques for network services verification.
- Fourth, we identify future research directions for verification approaches to detect anomalies.

**Outline**. We structure the rest of the paper as follows. Section II provides a background on the NFV architecture and the NFV threats and vulnerabilities that may cause network services anomalies. The section also proposes a generic definition of a network service anomaly. We present in Section III a taxonomy of network services' anomalies and analyze their impact on service attributes. In Section IV, we study existing techniques and approaches to verify anomalies. We finally discuss future research directions for anomalies verification in Section V and conclude this paper in Section VI.

## II. BACKGROUND.

### A. NFV Architecture

Network Functions Virtualization (NFV) advocates a network architecture using cutting-edge virtualization technologies to deliver network services in software instances independently of the underlying hardware. Fig .1 depicts the NFV reference framework's hierarchical model [1] defined by ETSI (European Telecommunications Standards Institute). The framework consists of three distinct layers: (1) the NFV Infrastructure layer (NFVI), (2) the VNFs layer, and (3) the NFV Management and Orchestration layer (NFV MANO).

The NFV Infrastructure includes physical servers, storage devices, and networking devices. Hypervisors, e.g., Xen, VMWare, KVM, partition these physical resources into logical resources to create VMs and virtual links that connect VMs. Although traditional hypervisors provide strong isolation between VMs, they pose performance issues [110] (e.g., significant VMs' boot and migration time) that prompt cloud providers to replace VMs with containers [109].

VNFs implement software versions of physical network functions such as WAN optimizers, firewalls, proxies, which, chained together, provide the network with new value-added functionalities, called network services.

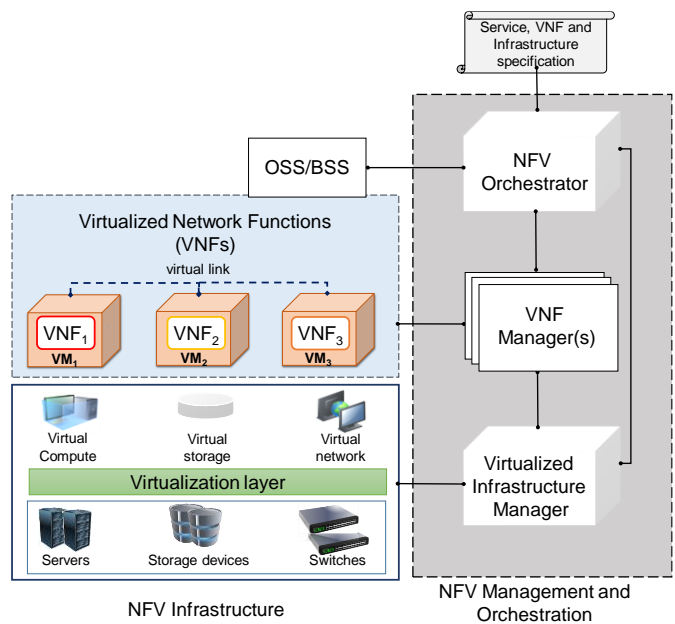

*Fig. 1. High-level layers of NFV reference framework.*

These virtual appliances run in VMs to facilitate their deployment, isolate their execution, and increase their scalability.

The NFV MANO manages and orchestrates the lifecycle of network services, VNFs, and resources. It comprises three functional components: (1) Virtualized Infrastructure Manager (VIM), (2) VNF Managers, and (3) NFV Orchestrator. The VIM manages and controls the NFV Infrastructure by ensuring resource allocation and deallocation to VNFs. A VNF Manager manages the lifecycle and the configuration of individual VNFs instances. The NFV framework may comprise multiple VNF Managers that may belong to different software vendors. Also, a single VNF Manager may serve a single VNF or a collection of VNFs. The NFV Orchestrator deploys the network service and manages the life cycle of network services. Tenants express their requirements to cloud providers with network services specifications, which dictates NFV Orchestrator's deployment and management decisions, including the amount and characteristics of resources to allocate to VNFs, the routing configuration to apply to virtual networks, the events to consider to scale VNFs, the constraints to consider to associate VNFs to physical servers, and the QoS requirements to apply to VNFs. The Operations and Business Support Systems (OSS/BSS) is a collection of applications helping cloud providers to manage their network and automate business functions such as service request management and billing. OSS/BSS is optional to the NFV architecture.

### B. NFV Attack Surface

The NFV framework depends on many heterogeneous software technologies organized in different layers. The NFV-specific threats and vulnerabilities lie at the intersection of both generic virtualization and traditional networking threats and vulnerabilities, e.g., side-channel attacks, memory leakage, co-residency attacks, flooding attacks. As a result, the framework exposes a more extensive attack surface than conventional networks. In this context, Pattaranantakul *et al.* [13] examined multiple threats related to the NFV architecture. They proposed a taxonomy that comprises threats to the NFV MANO, NFV Infrastructure, and VNFs layer. We summarize this taxonomy as follows:

- *NFV MANO layer's threats*: NFV MANO exposes vulnerabilities that an adversary could exploit to compromise network services. A trusted insider with privileged access to the NFV MANO's software can maliciously reconfigure the deployed network services. Adversaries can also exploit vulnerabilities, e.g., CVE-2018-15402, CVE-2019-1946, CVE-2020-3478, *CVE-2019-1971* in the external interfaces of the NFV MANO software to control the deployed services. Furthermore, synchronization issues or inconsistent configuration in the NFV MANO software could lead to inconsistencies between the network services and their specification.
- *NFV Infrastructure layer's threats*: the NFV Infrastructure threats range from traditional networks to virtualization threats. These threats include security issues in guest VMs, hypervisors, management interfaces, virtual resources, and hardware attacks such as side-channel attacks. Other threats specific to NFV environments include inadequate enforcement of security policies, multi-tenancy-related threats, and malicious insiders.
- *VNFs layer's threats*: the VNFs layer presents threats related to security management. Adversaries could take advantage of vulnerabilities in the VNFs software, e.g., CVE-2012-2663, CVE-2006-5276, to control the VNFs or violate security policies defined by tenants. Moreover, the menace often stems from the default configuration applied to VNFs or the insecure implementation of communication protocols, e.g., SSL and TLS. A malicious insider with sufficient permissions can get access to the VNFs and exfiltrates sensitive data. An adversary can also exploit vulnerabilities in the management interfaces to compromise the VNFs or violate users' privacy.

### C. Network services anomalies

A network service augments the network's value-added with functionalities that result from combining the distinct behavior of multiple VNFs. Tenants use a specification to define the network service's expected behavior throughout its lifecycle (instantiation, update, scaling, termination). The specification consists of clauses that individually contribute to different aspects of the network service's global behavior. Examples of clauses include network service's topology, network forwarding graph, resource allocation constraints, isolation policy, and scaling policy. We call *network service anomaly* a violation of at least one specification clause. We use the terms *anomaly*, *violation,* and *breach* interchangeably. Network services anomalies are the consequences of threats discussed in Subsection B. For example, a malicious insider with sufficient permissions could migrate VNFs to territories that enforce lax privacy' regulations, thus violating resource allocation constraints.

### III. TAXONOMY OF NETWORK SERVICE ANOMALIES IN NFV

Fig. 2 illustrates our proposed taxonomy of NFV-based network service anomalies, potentially leading to security and Quality of Service (QoS) issues. We describe six categories of anomalies related to six types of clauses violation: (1) Topology, (2) Forwarding graph, (3) VNFs, (4) Traffic filtering, (5) SLA, and (6) Ressources allocation. Simultaneously, we identify and analyze possible detrimental impacts of anomalies on critical service attributes: confidentiality, integrity, availability, performance, and resiliency. In the following, the term *adversary* designates an actor, either human or software, that can provoke a network service anomaly, intentionally or not.

### A. Topology anomalies

A VNF-level topology captures the structure of the network service by describing its constituent VNFs and virtual links that connect them. Two VNFs connected to two different virtual links cannot exchange packets directly. Thus, a topology specification can enforce a traffic isolation policy. As shown in Fig. 3, we identify four topologies anomalies [115], [116] that affect security and QoS:

- **Unexpected virtual links**: this anomaly occurs when an adversary creates a virtual link between two VNFs forbidden to communicate under a traffic isolation policy. The adversary can leverage the created virtual link to send legitimate or malicious traffic to another VNF. For example, the unexpected virtual link can serve to flood another VNF with denial of service packets.
- **Missing virtual links**: this anomaly corresponds to a network service instance missing one or more virtual links specified in the topology. We identify two possible consequences of this anomaly. First, the connection ceases between VNFs that communicate through the missing virtual links, causing service interruptions. Second, a sophisticated attack can consist of introducing this anomaly to force some VNFs to use another virtual link that has been priorly compromised. Then, the adversary can eavesdrop or tamper with the traffic traversing the compromised virtual link.
- *Missing VNFs*: like the previous anomaly, this anomaly occurs when one or more VNFs specified in the topology are missing in the network service instance. Consequently, some traffic escapes the processing by the missing VNFs, leading to violation of service objectives, depending on the missing VNFs' functions.
- *Unexpected VNFs*: an adversary can insert unspecified VNF into the topology. The inserted VNF can serve as a sniffer that extracts secrets from the traffic. The adversary can also alter the traffic arriving at the inserted VNF.

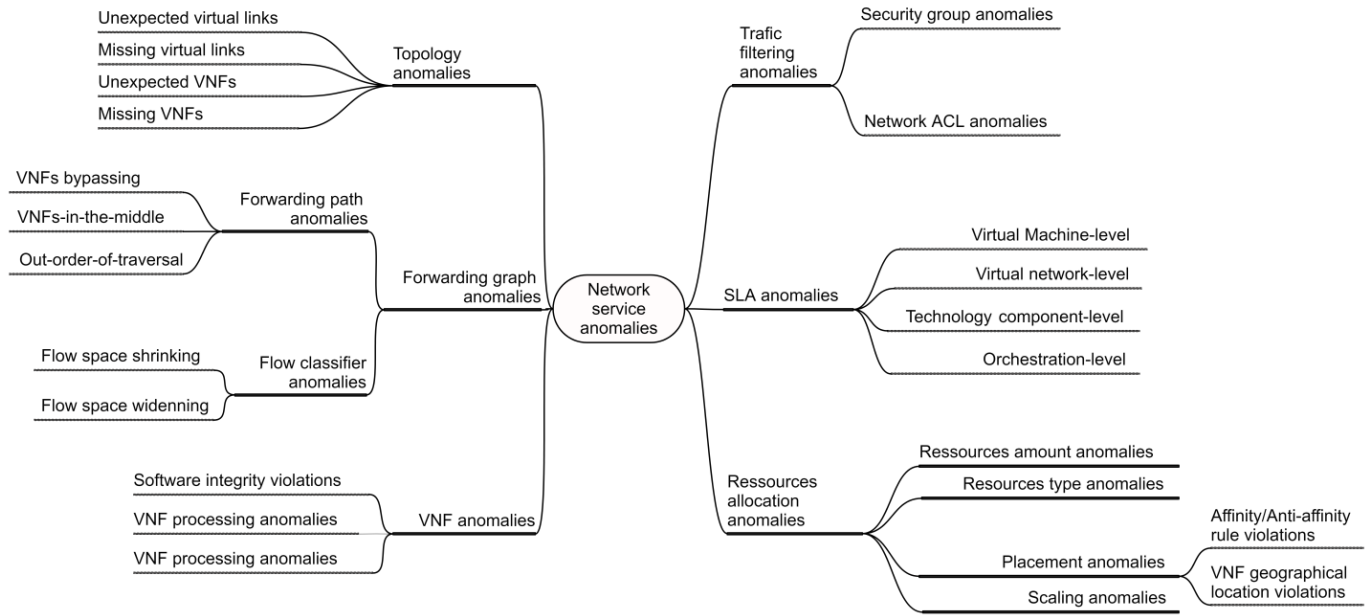

Fig. 2. Taxonomy of NFV-based network service anomalies. A node with a solid line represents either a class or subclass of anomalies. A node with a dotted line represents an example anomaly in a class or a subclass of anomalies.

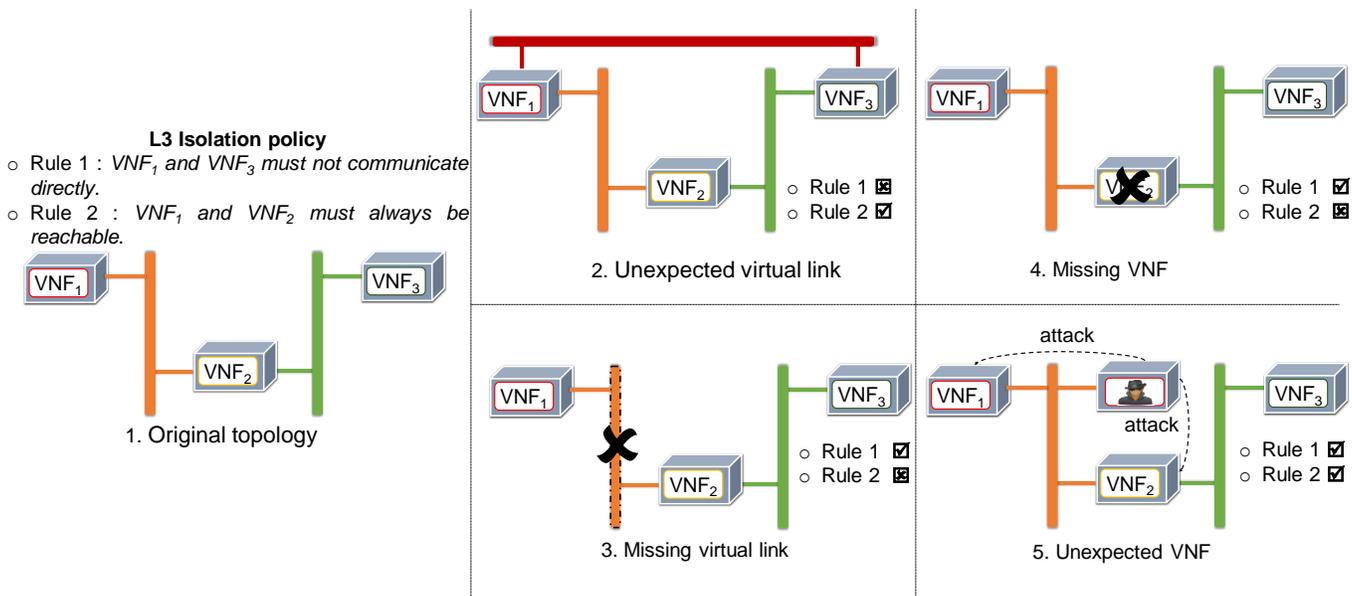

Fig. 3. Scenarios illustrating four topology anomalies. Solid bars represent virtual links. Thin bars connect virtual links to VNFs.

### B. Forwarding graph anomalies

A forwarding graph comprises two logical parts that together specify a traffic steering policy between VNFs : (1) *forwarding paths* which are ordered lists of VNFs that process a specific traffic class, and (2) *flow classifiers* that define traffic classes with classification rules, e.g., source IP, destination IP, port numbers. We then categorize forwarding graph anomalies into two subclasses : (1) *forwarding path anomalies* and (2) *flow classifiers anomalies*. Unlike *forwarding path anomalies*, which have been intensively explored in both non-NFV [30] and NFV [36] contexts, *flow classifiers anomalies* have been unreported in the literature. This latter anomalies' subclass proceeds from our more in-depth analysis. We observed that as well as forwarding paths, classification rules may also be violated, a possibility that scholars had overlooked. For instance, an adversary that comprises the NFV Orchestrator may change a tenant's specified classification rules. Also, a cloud administrator may misconfigure the classification rules.

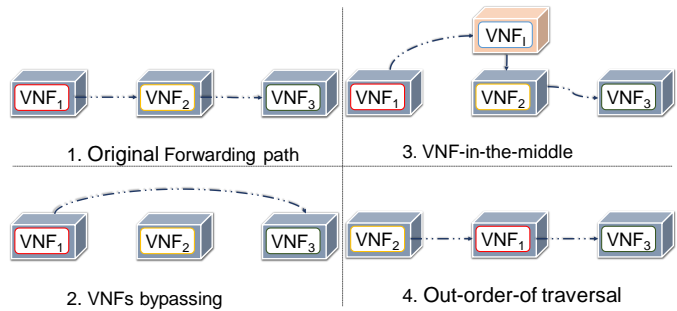

Fig. 4. Forwarding path anomalies.

*1) Forwarding path anomalies*

A forwarding path defines a list of VNFs that a specific traffic class must traverse in a particular order. Forwarding paths, also known as service function chains, allow differentiated processing of traffic classes based on their security and QoS requirements. A forwarding path anomaly occurs when an adversary violates one of the properties of a forwarding path, namely, the order, the number, or the type of VNFs. Fig. 4 depicts three types of forwarding paths anomalies that we identified :

- **VNFs bypassing**: we illustrate this anomaly through the following scenario. An adversary plans a denial-of-service (DoS) attack against a web service. However, before reaching the web service, each packet must traverse two VNFs: a load balancer and an Intrusion Prevention System (IPS). The IPS detects and mitigates DoS attacks' patterns. To bypass the IPS, the adversary gains control [14] over the switch connecting the VNFs. Then, the adversary reconfigures the compromised switch to forward packets outgoing from the load balancer to the web service instead of the IPS. Such a situation characterizes a VNF-bypassing anomaly. This anomaly can violate several service attributes, depending on the bypassed VNFs' functions, e.g., encryption, authentication, authorization, and traffic optimization.
- **VNF-in-the-middle**: an adversary can insert one or more VNFs in the head, the middle, or the tail of a forwarding path. As a result, the adversary can eavesdrop on end-users traffic. The inserted VNFs may also alter or extract secrets from the redirected traffic flows.
- **Out-order-of traversal**: this anomaly corresponds to a random permutation of the VNFs' order without inserting or bypassing some VNFs. This permutation distorts the expected behavior of the network service to random behavior. For instance, packets must first reach an IPS before being forwarded to an encryption proxy. Reversing this packets' processing order prevents the IPS from analyzing packets' payload, thus violating security objectives.

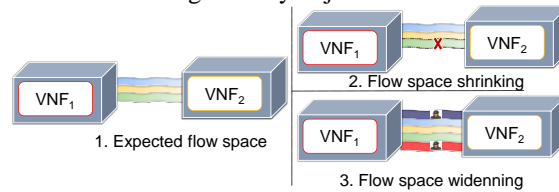

Fig. 5. Flow classifier anomalies

*2) Flow classifier anomalies*

Classification rules (e.g., IP address, TCP/UDP ports' number, L7 filtering) specify the traffic classes entering a forwarding path. The set of packets' headers matching these rules defines a geometric space [74], [85], as suggested by some researchers. A shrinking or a widening of this geometric space modify the traffic class authorized in a specific forwarding path. Thus, as shown in Fig. 5, we identify two flow classifier anomalies:

- **Flow space shrinking**: this anomaly results from an adversary modifying the classification rules to prevent legitimate traffic from entering a forwarding path. As a result, VNFs will illegitimately drop some traffic flows, making the end-service inaccessible to users.
- **Flow space widening**: the classification rules are modified to match unauthorized traffic flows in a forwarding path. For instance, an adversary can introduce this anomaly to authorize malicious traffic flows to enter a forwarding path handling only specific non-malicious traffic flows.

### C. VNF anomalies

The compliance of end-to-end network services depends on the correctness properties of individual VNFs in terms of software integrity [55]-[57], configuration, and execution [102]. We describe three types of VNF anomalies related to these properties as follow:

- **Software integrity violation**: VNFs are prepackaged as VM images that consist of a software stack, including guest OS, preinstalled applications, and libraries. A software integrity violation occurs when an adversary tampers with at least one of these software elements. Such violation happens in several scenarios. First, the adversary may replace a trusted library version with an older version that exhibits vulnerabilities. Second, the adversary may inject malware into a VM image to take control of the latter. Third, the adversary can exploit the NFV platform vulnerabilities to tamper with a VM during live migration [108] or at rest in the VMs images' database. Software integrity violations introduce security breaches in the VNF, alter its expected behavior, or degrades its performance.

- **VNF configuration anomalies:** configuration parameters control VNFs' expected behavior. For instance, the *iptables* rules' set defined by a tenant determines the incoming and outgoing traffic passing through a firewall. Furthermore, to configure a load-balancer with a weighted-based load balancing algorithm, a tenant assigns a weight to the backend servers according to their load-handling capacity. Let us consider $C$ as a configuration set producing a specific behavior $\Omega$ of a VNF. A configuration set $C'$ will produce a different behavior $\Omega'$. The VNF configuration anomaly happens when a tenant expects a VNF to be at configuration $C$ and the VNF is at configuration $C'$. This anomaly may impact the security and the performance of the tenant end-service. The previous examples show that a misconfigured firewall may allow unauthorized users to access or tamper with sensitive data. Misconfiguration in a load balance may result in imbalance traffic, thus increasing network latency. We highlight that such a type of violation has remained unidentified in the literature. This is most likely because configuration violations may easily be confused with software integrity violations. However, these two violations must be differentiated because, even if a VNF's software integrity remains unviolated, a change of its configuration will affect the network service performance and security.
- **VNF processing anomalies:** Each VNF implements a network function $f(p,C)$, which must return an expected output $O$ for each input packet p, given a configuration $C$. The expected output could be either a forwarding decision (drop, forward) or a new packet. A VNF processing anomaly [102] corresponds to situations where the function $f$ receives a packet p and returns an output $O'$ instead of $O$, whereas $C$ remains unchanged. This anomaly results from bugs, malware, or control flow attacks. Depending on the VNF holding the anomaly, the confidentiality, integrity, availability, and performance of the high-level service may be impacted. For example, a faulty IDS may not detect attack patterns such as SQL injection or cross-site scripting, leading to data confidentiality or integrity violation.

### D. Traffic filtering anomalies

Tenants define traffic filtering policy with security groups and network ACLs policies, each offering a particular defense layer. Whereas security groups apply to individual VNFs, network ACLs apply to all VNFs belonging to the same subnets[1]. Traffic filtering anomalies [74], [80] occur when an adversary authorizes a VNF to send or receive specific traffic, whereas a traffic filtering policy prohibits communication. Such an anomaly breaches the protection layer offered by traffic filtering, thus allowing malicious traffic patterns (e.g., DoS) to reach VNFs.

### E. SLA violations

According to ISO/IEC 20000-10:2018, a Service Level Agreement (SLA) is a contract that binds the cloud provider to provide network services for tenants at an agreed-upon level of performance. The cloud provider negotiates with each tenant some Key Quality Indicators (KQIs) [112]. Each KQI captures a critical aspect of the tenant's network service, i.e., VNFs outage downtime, VNFs reliability, VMs failure rate, VMs stall time, and service quality of life cycle management actions [113]. Tenants monitor each KQI by collecting and aggregating a set of Key Performance Indicators (KPIs) measurements, including minimum bandwidth, packet loss, and mean time between failures. Also, providers and tenants circumcise the bounds that KQIs/KPIs must meet. Each KQI/KPI's bounds define lower and upper warning thresholds and lower and upper error thresholds. For instance, a tenant may require a minimum bandwidth of 1G/s for all flows between VNFs. Additionally, the tenant may expect that given a corpus of 1000 flows between two specific VNFs, only one flow's latency can exceed 300 ms [15]. An SLA violation (a.k.a, SLA breach) [15], [23], [26] occurs when a KQI/KPI exceeds its specified bounds. In line with the KQIs' taxonomy proposed by ETSI [114], we distinguish four types of SLA violations, depending on the level at which the violation occurs:

- **VM-level SLA violations**: they correspond to violations of VM-level KQIs. An SLA may require some KQIs to assess the service quality of VMs, especially after their instantiation and their integration with VNFs. Such KQIs include VM stall, Premature VM Release Ration, VM scheduling Latency, or VM Clock Error [114].
- **Virtual network-level SLA violations**: these violations concern KQIs assuring the quality of virtual networks that connect VNFs and other network services' elements such as databases, storage systems, and physical network appliances. Examples of virtual network-level include packet loss, delay, and jitter.
- **Technology component-level SLA violations**: network services may rely on external technology components (i.e., databases-as-a-service, storages-as-a-service) offered by cloud providers in the platform-as-a-service model. KQIs such as service reliability, latency, and outage may be necessary to service to appraise the quality of these technology components. Technology component-level SLA violations happen when one or more of the latter KQIs exceeds its specified bounds.
- **Orchestration-level violations**: they concern the violations of KQIs that guarantee the quality of VMs and virtual networks' orchestration. These KQIs cover metrics such as VMs and virtual networks' provisioning latency and reliability, VMs Dead on Arrival, and virtual networks diversity compliance.

---

[1] We redirect the reader to reference [121] which provide an in-depth comparison between security groups and network ACLs.

SLA violations harm both the cloud provider's and tenants' business operations. When a cloud provider breaches an SLA, his reputation erodes toward the affected tenant, who may claim some compensations. The compensations may take the form of service credits or discounts for future billings. However, such remedies pale in contrast to the detrimental repercussions of SLA violations for the tenant, including lost customers and incomes, tarnished image, and reduced productivity.

### F. Resources allocation anomalies

*1) Resources amount anomalies:* The requirements for the amount of virtual resources such as CPUs, memory storage, and I/O network bandwidth vary from one VNF to another, depending on its functional characteristic and its expected level of QoS. A defect in resource amount [113] can occur in various ways when the amount of resources allocated to a VNF does not meet its specifications. This defect can result from errors when processing network service specifications. It can also be due to the service provider lying in the resource amount provided to a VNF. A lie close to the boundaries is hard to distinguish but still may affect the service performance.

*2) Resources type anomalies*

The optimal execution of certain VNFs requires resources with specific characteristics, such as CPU architecture, storage type (e.g., HDD, SSD), trusted execution environment. For example, privacy-preserving VNFs [107] often require processors with Intel SGX [47] support to execute. A resource type anomaly [113] occurs when a VNF instance runs with resources exhibiting different characteristics than specified.

*3) VNFs placement anomalies*

- *Affinity/Anti-affinity rule violation:* tenants control the placement of VNFs on physical hosts with affinity and anti-affinity rules. An affinity rule applies to a group of VNFs and requires their placement on the same hosts. Conversely, an anti-affinity rule spreads a group of VNFs across different hosts. Tenants commonly set the anti-affinity rules to achieve their high availability and resiliency goals. The violation of an affinity rule [114] happens when some VNFs belonging to the same affinity group are deployed on different hosts, thus adding latency in packets transfer between VNFs. Inversely, an anti-affinity rule violation [113], [114] occurs when some VNFs of the same anti-affinity group are executed on the same host. An anti-affinity rule violation compromises the high-availability properties because a host failure degrades the entire anti-affinity group's performance or causes the whole group's shutdown.

- *VNF geographical location violation:* network services specification allows tenants to constrain the geographical locations of hosts on which their VNFs execute. The geographical location constraint serves: 1) to meet the requirements of low-latency services by bringing VNFs closer to users, 2) to comply with regulations (e.g., GDPR [2], AAPs[3]) by hosting VNFs on jurisdictions adapted to users, or 3) to ensure fault-tolerance by dispersing VNFs across different availability zones. The VNF geographical location violation [56] occurs when an adversary places a VNF in a different geographical location than specified. Such anomaly induces performance and privacy concerns. For instance, end-users may experience significant latencies when they become distant from VNFs. Furthermore, when VNFs are relocated to territories with lax privacy regulations, current governments can arrogate the right to compromise end-users privacy.

*4) Scaling anomalies*

Tenants rely on a scaling policy to automatically adjust the number of instances of a VNF (horizontal scaling) or the resources allocated to a VNF (vertical scaling). Scaling policies define scaling events that trigger scaling operations[4]. A scaling event corresponds to a threshold violation of scaling metrics, including average CPU utilization and average response delay. A scaling anomaly [113] occurs when a scaling event happens without invoking the corresponding scaling operation.

### G. Summary of network service anomalies and their impact

From the previous taxonomy of network service anomalies, we summarize in TABLE *I* the possible negative impacts of service anomalies on service attributes, namely confidentiality, integrity, availability, performance, and resiliency.

---

[2] GDPR : General Data Protection regulation ; https://gdpr-info.eu/
[3] AAPs : Australian Privacy Principles ; https://www.oaic.gov.au/privacy/australian-privacy-principles/
[4] https://docs.aws.amazon.com/autoscaling/ec2/userguide/as-scaling-simple-step.htm

TABLE I
IMPACT OF NETWORK SERVICES ANOMALIES ON SECURITY AND QUALITY OF SERVICE OBJECTIVES.

| Network services anomalies | | | Confidentiality | Integrity | Availability | Performance | Resiliency |
|---|---|---|---|---|---|---|---|
| Forwarding graph anomalies | Topology anomalies | Unexpected virtual links | ✓ | ✓ | ✓ | | |
| | | Missing virtual links | ✓ | ✓ | ✓ | | |
| | | Missing VNFs | ✓ | ✓ | ✓ | ✓ | |
| | | Unexpected VNFs | ✓ | ✓ | ✓ | | |
| | Forwarding path anomalies | VNFs bypassing | ✓ | ✓ | ✓ | ✓ | |
| | | VNFs-in-the-middle | ✓ | ✓ | ✓ | | |
| | | Out-order-of traversal | ✓ | ✓ | ✓ | ✓ | |
| | Flow classifier anomalies | Flow space shrinking | | | ✓ | | |
| | | Flow space widening | ✓ | ✓ | ✓ | | |
| | VNFs anomalies | Software integrity violation | ✓ | ✓ | ✓ | ✓ | |
| | | VNFs configuration anomalies | ✓ | ✓ | ✓ | ✓ | |
| | | VNF processing anomalies | ✓ | ✓ | ✓ | ✓ | |
| Traffic filtering anomalies | | Security group violation | ✓ | ✓ | ✓ | | |
| | | Network ACLs violation | ✓ | ✓ | ✓ | | |
| | SLA anomalies | VM-level | | | | ✓ | |
| | | Virtual network-level | | | | ✓ | |
| | | Technology component-level | | | | ✓ | |
| | | Orchestration-level | | | | ✓ | |
| Ressource allocation anomalies | | Resources amount anomalies | | | ✓ | ✓ | |
| | | Resources type anomalies | ✓ | ✓ | | ✓ | |
| | Placement anomalies | Affinity/Anti-affinity violation | ✓ | ✓ | | ✓ | ✓ |
| | | VNFs geographical location violation | ✓ | ✓ | | ✓ | ✓ |
| | | Scaling anomalies | | | ✓ | ✓ | |

The symbol (✓) indicates that the anomaly can detrimentally impact the corresponding service attribute. Blank cells denote the opposite.

## IV. VERIFICATION OF NETWORK SERVICES IN NFV

Verification provides a tenant with an assurance that a specific clause (see Subsection II.C) in its network service specification' is being enforced by the cloud provider. This section scrutinizes existing verification techniques to detect network services anomalies.

### A. Selection criteria for verification techniques

*Organization.* We organize the verification techniques following our proposed network services' taxonomy. In other words, we categorize all verification techniques according to the clauses that they verify.

*Coverage.* We detected a partial or a total lack of verification techniques for several anomalies when conducting the literature review. Thus, for some anomalies, we also consider non-NFV specific verification techniques. With proper integration, these techniques can be applied to NFV.

*Quality.* At the first stage, we examined papers published in journals, conferences, workshops, and standardization organizations such as IETF (Internet Engineering Task Force) and ETSI. At the second stage, for the sake of concision, we focus on representative contributions for each verification field based on scientific soundness, completeness, and quality of evaluations. A quantitative and qualitative analysis of cited papers (see Table II) reveals that a significant cohort of references comes from journals and conferences with high-impact factors and rankings. As an illustration, consider *IEEE INFOCOM*, *USENIX NSDI*, and *ACM SIGCOMM Computer Communication Review*.

*Timeliness.* Most of the considered papers range from 2014 to 2021. The references include an insignificant number of old milestone papers that we consider relevant for unaddressed verification fields in NFV.

TABLE II
QUANTITATIVE AND QUALITATIVE CHARACTERISTICS OF THE REFERENCES SOURCES. THE SELECTED SAMPLE COMPRISES SOURCES COUNTING AT LEAST TWO PAPERS.

|  |  | Total citations' count in this paper | CORE rank (from core.edu.au) | h5 Index (from Google Scholar) | h5 Median (from Google Scholar) |
|---|---|---|---|---|---|
| Conference | ACM CoNEXT | 4 | A | 32 | 52 |
|  | ACM SIGCOMM | 6 | A* | 63 | 118 |
|  | CAV | 3 | A* | 37 | 52 |
|  | CloudCom | 2 | C | 21 | 42 |
|  | ESORICS | 2 | A | 31 | 51 |
|  | EuCNC | 2 | Not referenced | 25 | 52 |
|  | ICDCS | 2 | A | 45 | 77 |
|  | ICIN | 3 | National:France | 17 | 30 |
|  | IEEE INFOCOM | 7 | A* | 70 | 112 |
|  | IEEE NetSoft | 4 | B | 27 | 40 |
|  | NSS | 2 | B | 11 | 24 |
|  | USENIX NSDI | 14 | National:USA | 66 | 105 |
|  | USENIX Security | 2 | A* | 81 | 138 |
| Journal | ACM SIGCOMM CCR | 5 | Not referenced | N/A | N/A |
|  | IEEE/ACM Transaction on Networking | 3 | A* | 65 | 94 |
| Report | ETSI GS NFV | 2 | Not referenced | N/A | N/A |

The h5 index corresponds to the h index of articles published in the last five completed years. It corresponds to the highest value of h so that h articles published between 2016 and 2020 are cited at least h times each. The h5 median of a publication corresponds to the average number of times the articles comprising its h5 index are cited.

*B. Forwarding paths verification*

Researchers have proposed routing protocols [29]-[30] for forwarding path compliance verification in traditional IP networks. These approaches leverage essentially cryptographic techniques to validate the correct traversal of packets in a predefined forwarding path. ICING [30] stands out as a seminal contribution to forwarding path verification (***FPV***). ICING can verify whether a packet follows a pre-approved path on the network. The verification approach relies on Proof of Consent (PoC) and Proof of Provenance (PoP) to check path compliance. ICING first retrieves a PoC from each node on the path. PoCs are cryptographic tokens created by consent servers. PoCs certify that nodes consent to carry the packet along the path. PoPs include constructed Message Authentication Codes (MACs) and verify that the upstream nodes have handled a packet. ICING can defend against malicious and byzantine behaviors of consent servers, nodes, and providers.

However, traditional *FPV'* protocols do not apply to the NFV setting. Their underlying assumptions become unrealistic in an NFV context [36]. First, traditional protocols assume that forwarding paths are fixed and known in advance. In NFV, forwarding paths are dynamic and unpredictable. Second, traditional approaches suppose that network nodes are transparent to packets. However, middleboxes such as NATs and Proxies legitimately modify packets' headers and payloads. Such modifications exacerbate the difficulty of tracking packets along forwarding paths. Also, distinguishing between legitimate and illegitimate modifications becomes challenging [36]. For example, suppose an adversary compromise an HTTP proxy and injects a malicious code in the HTTP Body fields of packets. An unsophisticated verification protocol will mistakenly trust these modifications, as proxies are known to modify packets. Third, the traditional protocols assume a network forwarding behavior that only depends on flow tables. Nevertheless, NFV involves stateful VNFs whose forwarding decisions depend on VNFs' internal states. In other words, packets belonging to the same flow could be forwarded to different paths, depending on VNFs' internal states. Finally, traditional protocols cannot handle the complexity introduced by the highly dynamic nature of NFV, which offers migration and autoscaling capabilities. For instance, the scale-out operation requires a consistent per-flow or cross-flows states' synchronization [31],[32] between replicas to distribute traffic between these replicas transparently.

Recent works [33]-[41] have tackled forwarding paths' verification considering the new issues introduced by NFV.

Seyed Kaveh Fayazbakhsh *et al.* [35] analyzed the problems introduced by stateful and dynamic middleboxes actions (e.g.,

packets modifications, opaque behaviors). They show that middleboxes' actions could introduce unforeseen forwarding paths, leading to misdetection of forwarding path anomalies. They also illustrate how middleboxes violate the *origin binding* (i.e., binding between packets and their senders), a requirement for *FPV*. To address these problems, the authors advocate that middleboxes generate and add tags to outgoing packets. Downstream middleboxes consume tags to bind packets with their origin, whereas switches consider tags to steer packets across middleboxes. However, naïve packets' tagging becomes ineffective when one or more switches are compromised. For instance, a compromised switch may tag packets without forwarding them to the correct next-hop.

Thus, Kai Bu *et al.* [38] demonstrated that an attacker could perform a *middlebox-bypass attack* (i.e., VNFs bypassing anomalies) that resists traditional tagging and statistics-based verification. The authors suggest a sophisticated packets' tagging that consists of randomizing tags' generation, making them probabilistically unknown by compromised switches. The authors leverage this latter tagging technique to design FlowCoak, a real-time verification protocol that detects and prevents the *middlebox-bypass attack*.

Setting FPV' mechanisms may require modifying the internal logic of VNFs or the cloud framework[111], thus leading to complex and expensive development and deployment. To deal with such a challenge, Xiaoli Zhang [34] presented vSFC, a real-time FPV' scheme that exempts developers from modifying existing VNFs. Their verification scheme relies on a verification layer decoupled from VNFs' processing and embedded in VMs supporting VNFs. Each packet incoming or outgoing from VNFs must pass through the verification layer, which comprises an input module and an output module. Output modules tag packets outgoing from VNFs. Then, downstream VNFs' input modules verify incoming packets' tags to detect various forwarding path anomalies. As vSFC separates the verification layer from VNFs' logic, input and output modules can be deployed on trusted execution environments. In that direction, cSFC [37] proposes to execute both VNFs' processing and verification in SGX enclaves to preserve data confidentiality and VNFs software integrity against powerful adversaries.

Auditing the NFV and its compliance with the forwarding path was presented in AuditBox [36]. AuditBox offers a system for runtime guarantees on the compliance with forwarding path policies continuously. It supports dynamic forwarding path policies, packet modification, stateful behaviors of VNFs, and requires minimal modifications of VNFs. The verification system adopts a hop-by-hop attestation protocol to support dynamic forwarding path verification and reduce packet overheads. AuditBox trusts packets modifications by running each VNF's entire code in SGX enclaves, which offer hardware-based integrity protection. This latter property ensures that packets' modifications spring from uncorrupted VNFs. However, the proposed technique imposes significant overhead, likely induced by the hardware bottlenecks of Intel SGX. Running only the verification code in enclaves [34] could limit the overhead induced by Intel SGX.

In a published work in progress [120], the IETF has suggested a packets' *proof-of-transit* (POT) protocol in the context of service function chaining and traffic engineering. The protocol implements a *Shamir's Secret Sharing scheme* to cryptographically verify that packed has traversed all nodes on a specific path. However, the protocol overlooks essential aspects of NFV, including the unpredictability of paths. Also, the current version of the verification algorithm revealed some weaknesses, such as an inability to detect *stealth nodes*. Aguado et al. [40] employed a similar on top of the *Madrid Quantum Network*. The authors suggest many improvements in the IETF POT protocol, e.g., encrypting POT metadata in packets and reducing the protocol overhead. They leverage a quantum key distribution (QKD) system to secure the secret keys used to encrypt POT metadata.

Other works [33], [39] advocate static *FPV*. Specifically, Brendan Tschaen *et al.* [33] propose SFC-Checker, a static analysis-based framework, to check the correct forwarding behavior of dynamic and stateful forwarding paths. With a snapshot of the network state that includes the topology, flow tables, and the VNFs model, SFC-Checker creates a stateful forwarding graph using a Finite State Machine. SFC-Checker uses a static verification algorithm to troubleshoot and diagnose the forwarding behavior of forwarding paths.

Fulvio Valenza *et al.* [39] introduce a formal model that allows network administrators to specify forwarding policies and a broad spectrum of anomalies in a highly flexible and extendible manner. The formal model enables the detection of forwarding anomalies before their enforcement, thereby avoiding the wastage of resources required for translating and deploying anomalous forwarding policies.

In Table III, we present a comparison of representative works on forwarding path verification.

TABLE III
COMPARISON OF REPRESENTATIVE WORKS ON FORWARDING PATH VERIFICATION

| References | Legitimate packets modification | Unpredictable packets path | Stateful VNFs | Stateless | Verification chronology | Year |
|---|---|---|---|---|---|---|
| ICING [30] | | | | | realtime | 2011 |
| OPT [29] | | | | ✓ | real-time | 2014 |
| FlowTags [35] | ✓ | ✓ | ✓ | | real-time | 2014 |

| vSFC [34] | ✓ | ✓ | ✓ | | real-time | 2017 |
| --- | --- | --- | --- | --- | --- | --- |
| FlowCoak [38] | ✓ | ✓ | ✓ | ✓ | real-time | 2018 |
| AuditBox [36] | ✓ | ✓ | ✓ | | real-time | 2021 |
| SFC-Checker [33] | ✓ | ✓ | ✓ | | static | 2016 |
| Fulvio Valenza et al. [39] | | ✓ | N/A | N/A | static (Specification phase) | 2019 |

### C. Software package integrity verification

The software integrity of VNFs emerges as the most critical service attribute. When software becomes corrupted, the software functions that ensure compliance with other attributes (e.g., confidentiality, availability, and performance) may be compromised [55]. Thus, software packages' integrity verification (*SPIV*) of VNFs becomes crucial in outsourcing network services to the cloud.

This subsection surveys existing works on software package integrity verification. We identify two primary techniques for *SPIV*: digital signature and Trusted Computing. In TABLE IV, we provide a comparison of representative works leveraging these techniques.

Digital signature guarantees the integrity and authenticity of a piece of data. Digital signature-based *SPIVs* [56], [57] consist of generating trusted signatures of VNFs images and checking the integrity of each VNF against these signatures. OpenStack [57] allows tenants to leverage digital signatures and certificates to validate images before storage or download from the image database (Glance). Thus, this validation prevents tenants from storing or downloading compromised images from Glance. Additionally, Shankar Lal et al. [56] leverage a signing authority and a verification authority to check VNFs images' integrity. The signing authority priorly generates SHA256 digests of the VNFs' images and then signs them with its private key. When the verification authority receives a request with the fresh digest of a VNF's image as input, it uses the signing authority's root certificate to verify the image signature. However, an adversary that gains access [44], [45] to the trusted signatures base can manipulate some records to match them with signatures of altered VNFs images, thus bypassing the *SPIV*. Furthermore, an adversary that tampers with the cryptographic functions generating signatures can bypass the *SPIV* protocol.

To deal with the previous issues, the Trusted Computing Group [58] advocates leveraging Harward-based Root-of-Trust (RoT) to measure and store signatures or hashes of software (a.k.a *Prover*) to be verified, making these measurements immutable. Well-known Harward-based RoTs include Trusted Platform Module (TPM) [46] or Intel Software Guard Extension (SGX) [47], ARM's TrustZone [48]. These RoTs support a remote attestation [42] protocol in which an unmodified *Prover* convinces a remote *Verifier* that the former is in a trusted state. Upon receiving an attestation request, the RoTs computes a digital signature of the *Prover* that the latter uses as a token to authenticate itself to the *Verifier*. The *Verifier* relies on a trusted signatures database to appraise the trust level of a received token. Several works leverage Trust Computing technologies for *SPIV* [43], [50]-[52].

Cloud infrastructures comprise a large variety of physical hosts exhibiting different characteristics. Thus, remote attestation-based *SPIV* must ensure the flexibility of verifying the integrity of VNFs running on hosts supporting heterogeneous hardware RoTs.

To address such a challenge, Trust Monitor (TM) [50] leverages attestation drivers as plugins to implement remote attestation workflows for various RoTs, including TPM, Intel SGX, and AMD SEV. Furthermore, TM's framework integrates within the NFV MANO as a stand-alone module. In other words, TM separates the trusted monitoring and reporting procedures from standardized NFV MANO's workflows.

Other works [51]-[53] focus on integrity verification in the context of lightweight virtualization frameworks such as Docker. Specifically, the DIVE [53] framework leverages a modified version of the Linux Integrity Measurements Architecture (IMA) [54] to provide runtime integrity evidence of not only running containers but also the host and the container engine. When DIVE detects a specific container's compromise, it can terminate the latter or replace it with a new one. Although DIVE authors conduct their work outside of NFV, cloud providers could integrate DIVE into the NFV MANO to verify the integrity of containerized network functions (CNFs).

However, IMA exposes the internal states of a given tenant's containers to its co-tenants. The standard IMA protocol encloses the states of all containers hosted on the same server into a single measurements log (ML). During the verification protocol, each *Verifier* retrieves the entire ML. Hence, in a multi-tenant cloud, an adversary with access to ML can steal information on other co-hosted containers' internal states. From that information, the adversary can infer and then exploit the vulnerabilities that co-hosted containers exhibit. Thus, later works [51], [52] have considered privacy-preservation in IMA-based containers verification.

Container-IMA [51] partitions ML into virtual MLs called cPCRs (containers' Platform Configuration Registers). By parsing containers namespace, Container-IMA links them to cPCRs in a one-to-one association. To guarantee cPCRs' protection, Container-IMA binds them to Harward-based RoT such as TPM. Furthermore, unlike DIVE, Container-IMA can measure and verify the integrity of container dependencies (e.g.. libraries, files) and boot time (e.g., images and boot configurations).

Both DIVE and Container-IMA only focus on the integrity verification of individual containers. CloudVaults [52] goes beyond that by considering the integrity verification of the entire Service Graph Chain (SGC) of microservices-based applications while preserving privacy. CloudVaults tags an SGC as trusted if and only if all containers composing the SGC are correctly attested.

We present in Table IV a comparison of representative works on software package integrity verification.

TABLE IV

COMPARISON OF REPRESENTATIVE WORKS ON SOFTWARE PACKAGE INTEGRITY VERIFICATION.

| References | Techniques | Supported virtualization technologies | Hardware Root-of-Trust (RoTs) | | Scalability | Privacy-preserving: isolation of tenants' measurements | Service Graph Chain Verification | NFV integration |
|---|---|---|---|---|---|---|---|---|
| | | | Tested RoTs | Heterogeneity support | | | | |
| OpenStack [57] | Digital signature | VM+Container | | | N/A | N/A | | |
| Shankar Lal et al. [56] | Digital signature | VM+Container | | | N/A | N/A | | ✓ |
| TM [50] | Remote attestation | VM+Container | TPM | ✓ | | | | ✓ |
| DIVE [53] | Remote attestation | Container | TPM | | | | | |
| Container-IMA [51] | Remote attestation | Container | TPM | | ✓ | ✓ | | |
| CloudVaults [52] | Remote attestation | Container | TPM | | ✓ | ✓ | ✓ | ✓ |

### D. Traffic filtering enforcement verification

The traffic filtering enforcement verification (*TFEV*) could be reformulated as a reachability problem that determines which packets can be exchanged between two hosts [72], and by extension, between two VNFs. Thus, when a tenant policy prohibits two hosts or VNFs from communicating, the *TFEV* algorithm verifies that the latter are unreachable. There are two main approaches for analyzing reachability in a network: static analysis and dynamic analysis.

In general, static analysis techniques [71]-[85] operate on a snapshot of the configuration state of networking devices, including switches, routers, packet filters (e.g., firewalls), and packet transformers (e.g., NATs, proxies). The collected configuration state serves as input to generate a unified network state. Thus, the reachability can be analyzed using a formal method. Static analysis approaches differ from the formal method, e.g., SAT Solver [73], Binary Decision Diagram [79], SMT [81], symbolic execution [74],[85], used to model and reason on the network state.

Precisely, Anteater [73] models the network state as a set of boolean functions using the network topology and the forwarding tables of various network devices, e.g., firewalls, routers, switches. Network operators specify a wide range of network invariants such as isolation and loop-free forwarding that a SAT solver can verify.

HSA [74] leverages packet header bits to model packets as points in an *L*-dimensional geometric space where *L* is the maximum header length. HSA models the network's end-to-end behavior by composing transfer functions to capture various network devices' behavior. A transfer function transforms subspaces of the L-dimensional space to other subspaces. Finally, HSA computes reachability and checks slice isolation with algorithms based on algebraic operations, such as intersection, union, and difference on header spaces.

Tiros [80] and SecGuru [81],[82] are two industrial case-study leveraging automated theorem proving tools to provide a network reachability reasoning tool to AWS and Microsoft AZURE customers, respectively. Tiros builds a static model of the AWS network to check reachability properties. The model includes a specification that formalizes AWS components, e.g., subnets, NAT gateways, firewalls, load balancers, and a snapshot that describes the topology and network details. Tiros encodes the specification, the snapshot, and the reachability queries with the language of various reasoning engines such as the Datalog solver Soufflé [86], the SMT solver MonoSAT [87], and the first-order theorem prover Vampire [88]. Finally, Tiros leverages the solvers to answer the reachability questions. SecGuru [81],[82] automatically validates the correctness and consistency of network reachability policies in the Microsoft AZURE cloud. It encodes policies and semantic diffs with bit-vector logic formulas. SecGuru uses Z3, a Satisfiability Modulo Theories (SMT) solver, as engine analysis. With SecGuru, the AZURE operators can perform a regression test suite to proactively check policies before their deployment. The regression test suit avoids enforcing policies that may introduce security vulnerabilities or availability issues in the network.
Other works [85], [71], [89], [90] concentrate on monitoring the network configuration changes to detect change events that introduce security failures.

VeriFlow [85] continuously monitors the configuration in SDN networks to verify network invariants in real-time. It leverages a proxy that intercepts and checks the forwarding rules sent by the SDN controller to the network devices. VeriFlow optimizes the verification time with equivalence classes.

Cloud Radar [71] dynamically monitors virtualized infrastructures' configuration changes to detect near-real-time security failures (including network isolation) related to the topology. Cloud Radar represents the virtualized infrastructures with a graph model called the *Realization model*. It updates the *Realization model* with information retrieved from parsing change events. On the other hand, Cloud Radar expresses security policies such as network isolation, storage isolation, and VM placement as attack states using the graph model. To detect security violations, Cloud Radar tries to match the attack states with the *Realization model*.

Unlike static analysis, dynamic analysis techniques [91]-[96] generate and inject probing packets in the network to detect reachability issues. *Ping* and *Traceroute* are two primary tools that network administrators use for dynamic analysis. A key challenge in the dynamic analysis consists of generating a minimal set of probing packets that fully cover the testing of the isolation policy.

For the online checking of network reachability, ATPG [91] automatically generates a minimal set of test packets. ATPG uses the header space geometric framework (the same used by HSA [74]) to model the network state collected from various sources such as forwarding tables, ACL, and configuration files. Using header space analysis to find reachability between a set of test terminals, ATPG generates the minimal set of packets required to test every forwarding rule. Periodically, the test terminals send test packets in the network and use a fault localization algorithm to locate the cause of errors.

Monocle [93] checks the data plane correspondence with the SDN controller's intend network state in the presence of hardware and software failures and bugs. It adopts active monitoring to detect failed rules and links within a few seconds. Monocle leverages a proxy that intercepts all the rule modifications issued to a specific switch to maintain its expected content. Monocle then uses the expected state to generate probing packets to test each expected rule on the switch. The probing packets are generated by formulating the rules as Boolean Satisfiability problem. By injecting the probing packets in the network and observing how the switch modifies them, Monocle checks correspondence between the SDN controller view and the switch behavior.

Pronton [94] has a similar approach to Monocle but focuses on optimizing the probing packets' generation time and the number of generated probing packets. Pronto leverages the atomic predicate concept to determine the set of rules tested by a probe. Thus, Pronto can generate all the probing packets in a few seconds. Furthermore, conversely to prior works like ATPG and Monocle, Pronto uses one probe to simultaneously test multiple rules, reducing the number of generated probes.

*E. SLA compliance verification*

SLA compliance verification *(SLA CV)* detects SLA violations by determining whether the SLA performance and availability metrics are within the specified bounds [16]. Even before the advent of NFV, *SLA CV* has received significant attention from the literature [16]-[22]. Two fundamental approaches for *SLA CV* are active measurements [16]-[19] and passive modeling [20]. Active measurements involve a periodic sent of probe packets over the network to collect measurements, such as delay and bandwidth on the network state. Active measurements generate additional network traffic and only detect potential SLA violations once they occur. Conversely, passive modeling detects SLA violations before their occurrence. Passive modeling consists of parsing the network configuration to generate a quantitative model of the network. Thus, passive modeling can identify misconfiguration threatening the achievements of SLA goals. However, passive modeling fails to capture the dynamic traffic changes, thus missing some SLA violations.

With new emerging NFV use cases such as network services outsourcing and NFV-based network slicing, SLA CV has gained greater importance. Only a few publications have tackled SLA CV in the context of NFV. SLA-Verifier [15] is the first SLA verification system to assess SLA performance metrics' compliance, such as latency, hop count, and network availability in NFV environments. The authors introduce a quantitative model of the network to perform static verification. However, static verification can fail to detect some SLA violations due to traffic' dynamic changes. To handle this issue, SLA-Verifier additionally performs online measurements. Exploiting heuristics algorithms, they select the optimal measurement type (passive measurement, active measurement) based on static verification results. Xiaoli Zhang and al.[23] proposed a performance compliance verification scheme for stateful middleboxes outsourced in an untrusted cloud which may deliberately manipulate verification procedures. Their approach leverages delayed sampling and commitment techniques to defend against cloud potential cheating behaviors. Jaafar Bendriss and al. [24] presented a cognitive SLA enforcement framework in SDN/NFV networks. Their framework collects raw metrics from running VNFs and leverages Artificial Neural Network to predict future SLA violations. Other works [25]-[28] provide architectural principles for SLA management and verification in 5G networks. More specifically, Apostolos Papageorgiou and al.[25] designed a specific SLA management architecture and workflows for 5G network slicing. They defined dynamic formulas for the computation of 5G network slicing-specific SLA metrics.

TABLE V compares representative contributions to the *SLA CV* regarding the verification approach, the assumption of an untrusted environment, and predictability. The verification approach is either based on active measurements or passive modeling. An assumption of an untrusted environment means the verification scheme can resist cheating behaviors. Predictability assesses whether the SLA violations can be detected before they happen.

*F. VNFs' geographical location verification*

VNFs' geographical location' verification (**GLV**) verifies that the VMs allocated to VNFs are hosted in an expected geographical location. We observed that VNFs' *GLV* has not yet received sufficient attention from the literature. Except for one work [56], almost no work considers *GLV* in NFV. In contrast, researchers have extensively tackled *GLV* in traditional IP networks and non-NFV cloud computing' scenarios. The proposed *GLV* techniques can be recycled for *VNFs' GLV*. We describe the latter techniques in the following.

Pratical approaches for *GLV* encompass: distance-bounding protocols [59], landmark-based geolocation [63], [70], and IP address mapping based geolocation [64]-[67], topology-aware geolocation [68]-[69], and hardware-based geolocation [56], [60]-[62].

Distance bounding protocols [59] cryptographically verify the upper-bound distance of a *Prover* to a *Verifier*. The Verifier measures the round-trip time (RTT) between sending out challenge bits and receiving the prover's corresponding bits to estimate the Prover's location.

Landmark-based geolocation techniques [63], [70] assume a correlation or mapping between network metrics (e.g., Round Time Trip, hop counts, bandwidth, etc.) and geographic distance. To geolocate a target server, they consider a set of landmarks with known geographic locations. A landmark-based geolocation protocol first sends probing packets such as ICMP and HTTP packets between the target server and landmarks and measure the corresponding network metrics. The collected network metrics serve to generate a distance-to-delay function. The latter function predicts the distance between the target server and landmarks. The predicted distances are then used as inputs to triangulation techniques that allow the target server's geolocation. However, the distance-to-delay function's accuracy may be impacted [70] by network topology-related issues such as circuitous end-to-end paths.

Topology-aware geolocation techniques [68]-[69] enhance landmark-based geolocation algorithms' accuracy by geolocating intermediate routers between the target server and landmarks. *"Starting from the landmarks, the geolocation algorithm iteratively estimates the location of all intermediate routers on the path between the landmark and the target. This is done solely based on single-hop link delays, which are usually significantly less circuitous than multi-hop end-to-end paths, enabling topology-aware geolocation to be more resilient to circuitous network paths than delay-based geolocation."* [70]

Hardware-based geolocation [56], [60]-[62] approaches leverage tamper-proof hardware modules attached to the cloud server to guarantee its location. Hardware modules serve as roots-of-trust that store information on the server's geographical location. For instance, GeoProof [60] combines the proof of storage protocol and the distance-bounding protocol to provide geographic location assurance of data outsourced to a cloud. The GeoProof architecture involves a tamper-proof and GPS-enabled device attached to the cloud service provider's local network. The attached device is leveraged to run a distance-bounding protocol with the data centers. In [61], TPMs are placed on physical machines as a unique identifier. A third-party auditor maintains a trusted database of the location of each TPM. Shankar Lal *et al.* [56] present a proof-of-concept framework for biding VNFs to TPMs-enhanced hosts that satisfy geographical location constraints. They suggest including these geographical location constraints in VNFs' images' metadata and embedding the hosts' geographical information on their attached TPMs. TPMs attest to hosts' geographical location. VNFs placement involves matching them to hosts, which TPMs geolocalisation information matches VNFs' geographical location constraints.

TABLE V
*COMPARISON OF SLA COMPLIANCE VERIFICATION IN THE LITERATURE.*

| References | Active measurement (online) | Passive modeling (offline) | Assumption on untrusted environment | Predictability |
|---|---|---|---|---|
| SLA-Verifier[15] | ✓ | ✓ | | ✓ |
| Xiaoli Zhang and al.[23] | ✓ | | ✓ | |
| Jaafar Bendriss and al.[24] | ✓ | | | ✓ |
| 5GTANGO[26] | ✓ | | | ✓ |

TABLE VI
GAP ANALYSIS FOR NETWORK SERVICES' ANOMALIES VERIFICATION.

|  |  |  | Application to NFV | Traditional networks |
|---|---|---|---|---|
| Topology anomalies |  | Unexpected virtual links | ✓ |  |
|  |  | Missing virtual links | ✓ |  |
|  |  | Missing VNFs | ✓ |  |
|  |  | Unexpected VNFs | ✓ |  |
| Forwarding graph anomalies | Forwarding path anomalies | VNFs bypassing | ✓ | ✓ |
|  |  | VNFs-in-the-middle | ✓ | ✓ |
|  |  | Out-order-of traversal | ✓ | ✓ |
|  | Flow classifier anomalies | Flow space shrinking |  |  |
|  |  | Flow space widening |  |  |
| VNFs anomalies |  | Software integrity violation | ✓ | ✓ |
|  |  | VNFs configuration anomalies |  |  |
|  |  | VNF processing anomalies | ✓✗ | ✓ |
| Traffic filtering anomalies |  | Security group violation | ✓ | ✓ |
|  |  | Network ACLs violation | ✓ | ✓ |
| SLA anomalies |  | VM-level | ✓ | ✓ |
|  |  | Virtual network-level | ✓ | ✓ |
|  |  | Technology component-level | ✓ | ✓ |
|  |  | Orchestration-level | ✓ | ✓ |
| Ressource allocation anomalies |  | Resources amount anomalies |  |  |
|  |  | Resources type anomalies |  |  |
|  | Placement anomalies | Affinity/Anti-affinity violation |  |  |
|  |  | VNFs geographical location violation | ✓✗ | ✓ |
|  |  | Scaling anomalies |  |  |

The symbol (✓) denotes the existence of verification techniques in the corresponding context (NFV, traditional networks). The symbol (✓✗) denotes that the topic is relatively recent and tackled by very few works.

## V. GAP ANALYSIS AND FUTURE RESEARCH DIRECTIONS

As a recent technology, NFV remains in its maturation cycle. Its wide adoption requires addressing primary challenges, including security, trust, and compliance [97]. We expect network services' verification to ensure trust establishment in NFV deployment scenarios. However, the NFV literature has not fully invested in the necessary research activities in this area. In this section, we identity some research directions for network services' verification in NFV. Table VI provides a summary of the existence of verification techniques for each network service's anomaly. We distinguish traditional verifications techniques from the NFV-specific techniques because research efforts are needed to integrate traditional NFV architecture techniques. Even the tackled anomalies require further investigation. We focus on the following research tracks :

- **Cross-layer verification**. The main strength of NFV lies in its multi-layer architecture, which consists of both decoupling and independence of the NFV MANO's, the NFVI's, and the VNFs' layers. These properties simplify the management of network services and allow the automation of their life cycle. However, as some earlier studies [98], [117], [118] have shown, the multi-layer nature of NFV introduces inconsistencies issues between layers. For instance, consider a scenario where NFV Orchestrator sends configuration commands to other NFV components to deploy a network service. Suppose these components fail to apply these commands because of synchronization issues, bugs, software compromise, or failures. In that case, the network service state will be inconsistent across the layers. Consequently, verifying the network service at a single layer becomes skewed. Cross-layer verification is therefore crucial to operate network services in NFV environments. Such a verification approach may require details for mapping network services' states across the layers [98]. However, this mapping conflicts with the NFV philosophy that rests on the decoupling and independence of layers. Thus, a decisive question for the NFV community will be whether to enrich the NFV architecture with mappings details at the expense of the decoupling and independence of layers ? or to consider other research tracks to achieve the cross-layer verification? A study path to avoid mapping details consists of observing each layer's behavior to infer [119] network services' state at this layer.

- **Verification from tenant view.** When tenants mistrust a cloud provider, they cannot set their verification mechanisms by naively relying on the network services' states claimed by the cloud provider. A dishonest cloud provider can violate network specifications deliberately and conceal the violations by lying on the network services states. Thus, verifying network services from tenants' views is essential for the widespread adoption of NFV. Such a verification perspective remains challenging because tenants hold partial visibility over network services states. Tenants' visibility differs according to the cloud provider's transparency policy and its service model. For example, Software as a Service exposes only information on software instances. In contrast, Infrastructure as a Service exposes information on VMs instances. So a critical research question is *how to infer network services states from external and partial observability*?

- **Stateless forwarding graph verification**. NFV offers the benefits of deploying elastic, resilient, and highly available network services. VNFs replicas can be launched upon failure or to meet SLAs and availability requirements. VNFs can also be relocated to different points of presence to optimize network latency or allow hardware maintenance. Constant migrations and replications of VNFs require migrating states across VNFs instances while maintaining these states' consistency [99]-[101]. Thus, forwarding graph verification protocols should consider the migration and replication aspects of NFV. Designing a stateless protocol is a possible track to tackle these challenges.

- **Secure cryptographic keys**. Forwarding path verification protocols generally rely on symmetric keys shared among VNFs. During the network transit or at storage, these keys' security management is challenging in NFV because of its dynamic property (volatile VNFs, migrations, and replications). A single compromised key could break the security of the protocol. For instance, an attacker could escape the verification protocol by manipulating the verification procedures using the compromised key. Thus, protocols leveraging cryptographic hardware and procedures to protect the keys' transit and storage should be considered. For instance, recent work [40] from 2020 leveraged a Quantum Key Distribution - *the Madrid Quantum Network* - for highly secure provisioning of secret keys.

- **Flow classifiers verification.** To the best of our knowledge, we have noticed that flow classifiers anomalies verification has not been studied in the literature, despite the security issues they could introduce

- **VNFs processing anomalies**. One of the major drawbacks of service outsourcing is the loss of direct control and visibility over the correct execution of VNFs. Hence, there is no guarantee that VNFs are processing traffic payloads as expected. For instance, a compromised IDS could classify packets inconsistently with its signatures base, leading to undetected intrusions. Thus, research efforts must focus on mechanisms verifying the correctness of outsourced VNFs execution. Yuan *et al.* [102] laid the first stone in this domain by proposing a system that verifies string pattern-matching in untrusted cloud environments.

- **Tailoring SLA metrics to VNFs characteristics:** The SLA metrics must precisely define the tenants' expectations. However, traditional metrics such as bandwidth and latency are unsuitable in measuring the effectiveness of VNFs with different functional goals, ranging from network optimization to security functions. An Intrusion Detection System (IDS) security may be evaluated according to metrics such as detection rate and false alarms rate [105]. Simultaneously, a firewall evaluation may rely on its robustness to penetration tests [106]. Conversely, tenants may evaluate the performance of a load-balancer in terms of throughput and fault tolerance. In summary, the SLA metrics must suit the characteristics of each VNF. Thus, efficiently enforcing SLA in NFV requires a taxonomy on effectiveness evaluation metrics for VNFs, regarding their functional goals.

- **Automatic SLA compensation in NFV**: Services providers are legally committed to achieving the SLA goals to the extent possible. If a provider fails to meet service requirements, he must financially compensate tenants with service credits. However, submitting a service credit claim and receiving compensation is a complex and manual procedure that likely ends in disputes between providers and tenants [103]. Thus, it is crucial to integrate an automatic SLA compensation mechanism in the NFV framework. On the one hand, such automation involves both providers and tenants monitoring the deployed network services. On the other hand, both parties must trust each other's measurements. Smart contracts relying on secure digital ledgers such as Blockchain can help to establish the required trust. Several works [103], [104] have started this investment, but this track deserves more attention, especially in the context of NFV.

## VI. CONCLUSION

NFV's maturation has reached its critical stage. Its massive adoption in the industry requires building trust in NFV platforms and providers. We believe that *verification* will be a crucial feature for instilling confidence in enterprises to embrace NFV.

The present survey provided a state of knowledge on verification in NFV by introducing a taxonomy of possible network services' anomalies and reviewing the existing mechanisms to detect these anomalies. We motivated the importance of verifying network services anomalies by analyzing their detrimental impact on most critical service attributes, i.e., security, performance, and resiliency. Additionally, we examined the gap and challenges in achieving verification in NFV. We are confident that our survey will stimulate the production of abundant and relevant works on that topic. For instance, we observed that several anomalies remain unaddressed by the literature. Furthermore, many traditional verification systems should also be tailored to NFV because of its unique multi-layer and dynamic characteristics.

ACKNOWLEDGMENT

This study has been carried out with financial support from the French State, managed by the French National Research Agency (ANR) in the frame of the "Investments for the future" Programme IdEx Bordeaux - SysNum (ANR-10-IDEX-03-02).